\pdfoutput=1
\documentclass[11pt,twoside]{article}

\usepackage{asp2006}
\usepackage{graphicx}
\usepackage{lscape}

\begin{document}
\title{The elaboration of spiral galaxies: 
morpho-kinematics analyses of their progenitors with IMAGES}   
\author{F. Hammer, on behalf of the IMAGES collaboration}   
\affil{GEPI, Observatoire de Paris \& CNRS}    

\begin{abstract} 
The IMAGES (Intermediate MAss Galaxy Evolution Sequence) project aims at measuring the velocity fields of a representative sample of 100 massive galaxies at z=0.4-0.75, selected in the CDFS, the CFRS and the HDFS fields. It uses the world-unique mode of multiple integral field units of FLAMES/ GIRAFFE at VLT. The resolved-kinematics data allow us to sample the large scale motions at $\sim$ few kpc scale for each galaxy. They have been combined with the deepest HST/ACS, Spitzer (MIPS and IRAC) and VLT/FORS2 ever achieved observations. Most intermediate redshift galaxies show anomalous velocity fields: 6 Gyrs ago, half of the present day spirals were out of equilibrium and had peculiar morphologies. \\
The wealth of the data  in these fields allow us to modelize the physical processes in each galaxy with an accuracy almost similar to what is done in the local Universe. These detailed analyses reveal the importance of merger processes, including their remnant phases. Together with the large evolution of spiral properties, this points out the importance of disk survival and strengthens the disk rebuilding scenario. This suggests that the hierarchical scenario may apply to the elaboration of disk galaxies as it does for ellipticals.
\end{abstract}

\vspace*{-1.0cm}

\section{Introduction} 

Half of the present-day stellar mass density has been formed since z=1, during the last 8 Gyr (Dickinson et al. 2003; Fontana et al. 2003).  This result is particularly robust as it has been derived from two independent methods: the evolution of the cosmic stellar mass density and by integrating the universal star formation density including infra-red measurements (Flores et al, 1999). The two methods have their own weaknesses depending on the contribution of massive stars to the near-IR light (e.g. Maraston et al., 2006) or to the conversion of the mid-IR luminosity in star formation rate. However the good agreement between their predictions -at least below z=1- is rather compelling. 

Hammer et al. (2005) and Bell et al. (2005) have shown that most of the stellar mass formation during the last 8 Gyrs is associated to Luminous IR Galaxies (LIRGs, SFR$>$ 19 $M_{\odot}$) which may account from 50\% to 100\% of the star formation density in the z=0.5-1 range. Most LIRGs have stellar masses ranging from 2 to 20$\times$ 10$^{10} M_{\odot}$, those being responsible for the bulk of the star formation density reported by deep galaxy surveys (e.g. CFRS) and by studies of the past history of present-day galaxies (e.g. Heavens et al., 2004). Present-day intermediate mass galaxies are mostly spiral galaxies (70\% of them from the SDSS, Nakamura et al., 2004). How galaxies, mostly spirals with masses similar to that of the Milky Way, have assembled half of their stars over the past 8 Gyrs? To identify the main physical processes at the origin of the star formation, we are now embarked in a complete study of their progenitors, i.e. galaxies having emitted their light 4 to 7 Gyrs ago.

\vspace*{-0.5cm}
\section{One hundred distant galaxies with spatially resolved kinematics from the IMAGES survey} 

We gathered a sample of 100 galaxies selected on the sole basis of their absolute magnitude in J band ($M_J(AB)$$<$-20.3 corresponding to $M_{stellar}>$ 1.5 10$^{10} M_{\odot}$) and their redshift (0.4 $<$ z $<$ 0.9).  A considerable effort was made to ensure that our sample is representative of the intermediate mass galaxies at z$\sim$ 0.6. First, observations of their kinematics with VLT/GIRAFFE require the presence of the [OII]$\lambda$3726,3729 doublet in their spectra. Flores et al. (2006) and Yang et al. (2008a) convincingly demonstrated that for all the targets with $W_0([OII])>$ 15\AA,  sufficiently high S/N velocity fields may be retrieved after exposure times from 8 to 24 hours with VLT/GIRAFFE.  However, this does not account for galaxies without or with faint emission lines that represent 40\% of the galaxies at z$\sim$ 0.6 (e.g. Hammer et al., 1997). These galaxies are quiescent galaxies mostly made of E/S0 and quiescent spirals (e.g. Zheng et al. 2005; Delgado et al. 2009, in preparation).  In the following they will be considered as having relaxed kinematics, either supported by dispersion or by rotation. Second we have verified that our sample is representative of the galaxy luminosity function observed at z$\sim$ 0.6 (see e.g. Ravikumar et al, 2007). In the latter study it was shown that the luminosity density in UV, near-IR and mid-IR of our sample is indeed representative of the median value observed in the same redshift range. Third, galaxies were selected in 4 different fields of view (Yang et al. 2008) to minimise possible cosmological variance effects.Thus our sample is mostly limited by the Poisson statistical variance related to the number of galaxies with resolved kinematics. Notice that it is by far the much larger existing sample of distant galaxies with spatially resolved kinematics, because GIRAFFE at VLT is still the unique multi-IFU spectrograph allowing the observation of 15 galaxies at the same time.

An important step is the methodology used in identifying the nature of the velocity fields of distant galaxies. Because of their distances, the GIRAFFE IFU cannot resolve rotation curves as it can be done for local galaxies. This had let Flores et al. (2006) to propose a robust method to classify their velocity fields by using the supplementary information of their dispersion maps.  To illustrate this, let us consider a rotating disk. In its outskirts the dispersion map with 0.52 arcsec pixel ($\sim$ 3 kpc) is able to recover the random motions within the disk, while in the centre it samples the convolution of these motions with the large gradient of the rotational curve. This unavoidably results in a dispersion peak located at the mass centre, in the middle of the two extrema velocities. A diagnostic diagram thus tests the discrepancy of the dispersion peak both in location and intensity, through a comparison with expectations from the observed/modeled velocity field (see details in Flores et al., 2006; Yang et al., 2008a)\footnote{This has been allowed by the absence of cross talks in the GIRAFFE design.}. Notice also that the high spectral resolution of GIRAFFE (R$\sim$13000) ensures a proper removal of sky lines, the large exposure time warrants a very high S/N ($>$3 for each pixel, average of 10) and and the presence of the [OII]$\lambda$3726,3729 doublet provides reproducible measurements of the kinematics.

\begin{figure}[t]
\begin{center}
\includegraphics[scale = 0.5]{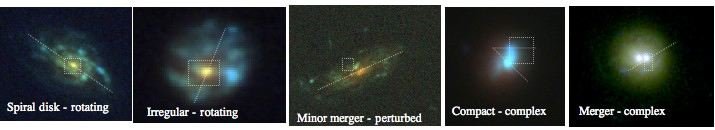}
\end{center}
\vspace*{-0.8cm}
\caption{Images of z$\sim$ 0.6 galaxies are combined b+v, i and z frames from HST/ACS. On each galaxy are superimposed the dynamical axis (dotted lines) and the dispersion peak (box with vertical size of 0.5 arcsecond). Morphological and kinematics classifications from Neichel et al (2008) and Yang et al (2008), respectively.   }\label{fig1}
\end{figure}

Three kinds of velocity fields were retrieved by Flores et al. (2006), including rotating disks (rotation axis following the main optical axis and dispersion peak in the centre), perturbed rotations (rotation axis following the main optical axis but offset of the dispersion peak) and complex kinematics (kinematical axis not aligned with the optical axis, or chaotic velocity fields and dispersion maps). Examples can be found in Flores et al. (2006), Puech et al. (2006), Yang et al. (2008a) and Figure 1. It results that among the 63 velocity fields studied in Yang et al.,  32\% are rotationally supported, 25\% are perturbed rotation, 43\% are complex. Notice that 9\% are too compact for being resolved with GIRAFFE and in the following, they are assumed to follow the same distribution than the above. Accounting for the whole population of z$\sim$ 0.6 galaxies, this reveals 33\% of rotating disks, while 41\% have anomalous kinematics, including 26\% with complex velocities. GIRAFFE is only sensitive to large scale motions and not to small variations as those caused by bars or by ordinary warps. Thus galaxy kinematics evolve strongly since the last 6 Gyrs (Yang et al., 2008a). Galaxies with anomalous kinematics are responsible for all the observed dispersion of the Tully-Fisher relation at z$\sim$ 0.6 (Flores et al. 2006; Puech et al. 2008a): it evidences how strong their kinematics are perturbed.

\vspace*{-0.5cm}
\section{Half of the present-day spirals had peculiar morphologies and anomalous kinematics, 6 Gyr ago}

The fields in which IMAGES galaxies have been studied all possess deep and high resolution imaging with two to four colours, mostly from deep exposures with HST/ACS. For a spiral galaxy, this ensures that we are able to identify galaxy morphological features within the optical radius up to z=0.5. Neichel et al. (2008) described a semi-automatic decision tree to classify the distant galaxy morphologies, based on the systematic use of the GALFIT software, of the calibrated and S/N weighted  colour maps  (Zheng et al., 2005) and finally visual inspections by three independent co-workers. We chose a very conservative method to classify galaxy morphology, keeping in mind the well known morphologies of local galaxies that populate the Hubble sequence. For example, we did not try to evaluate the morphology of compact galaxies (classified as compact), all galaxies for which the GALFIT sofware failed were classified as peculiar or merger, and we further imposed that spiral galaxies, whenever they possess a bulge, it should have a redder colour than the disk.

\begin{table}
\caption{Morpho-kinematical classification of  z$\sim$ 0.6 galaxies from Neichel et al.; for comparison, the last column shows the fractions derived from the SDSS (Nakamura et al; 2004) for galaxies in the same mass range. }
{\scriptsize
\begin{tabular}{lcccc} \hline
Type   & $z\sim 0.6$ & $z\sim 0.6$ & $z\sim$ 0.6   & local \\ 
       & $W_0(OII)\ge$15\AA~ & $W_0(OII)<$15\AA~ & all galaxies & galaxies \\
       & Neichel et al.  & Zheng et al.  & Neichel et al. & Nakamura et al. \\
       & & Delgado et al. & & Hammer et al. \\ \hline 
E/S0        &   0\% &  57\% & 23\% & 27\%\\
Rotating spiral   &  27\% & 43\%  &  33\% & 70\%\\
Peculiar/comp./merger  &  73\% & 0\% &  44\% & $\sim$ 3\%\\
  & & & & \\
With anomalous kinematics &  68\% & 0\%   & 41\%  &  \\ \hline
\end{tabular}
}
\end{table}

Applying this classification to emission line galaxies observed by GIRAFFE, we find only 29\% of spiral galaxies. We have thus compared our morphological classification to that of their kinematics, and found a remarkable agreement (e.g. Neichel et al., 2008). Almost all (95\%) but one galaxy with complex velocity fields have peculiar, compact or merger morphologies and most galaxies (80\%) with rotational velocity fields have spiral morphologies. Such an excellent agreement brings a considerable support to our conservative classification scheme. It should not be very sensitive to star formation since all the GALFIT measurements were done in the observed z band (rest-frame V band at z$\sim$ 0.7). In contrast, semi-automatic classification methods such as C-A or Gini-M20 are not predictive of their kinematics and strongly overestimate the number of spiral galaxies.

The combination of morphological and kinematical classifications results in a quite small fraction (16\%) of rotational spiral disks with emission lines ($W_0(OII)\ge$ 15\AA). Table 1 summarises the statistics at z$\sim$ 0.6 and compares them to local galaxies from SDSS (Nakamura et al.). We were conservative in doing such a comparison assuming that rotating spirals should have spiral morphologies and have a rotating velocity field.  Similar statistics combining kinematics and morphology does not exist for local galaxies, although this is in progress (Puech et al., 2009, in preparation).

Table 1 evidences that E/S0 were mostly in place at z$\sim$ 0.6, while half of rotating spirals were not. Six Gyrs ago, half of the local spirals had peculiar morphologies and anomalous kinematics. This result supersedes earlier results from Lilly et al. (1998) which were based on lower spatial resolution and S/N imagery, without kinematics. Most of the star formation is related to LIRGs that number density evolves also considerably (by factors 30 to 40, e.g. Elbaz and Cesarsky, 2003). Thus spiral galaxies are aggregating half of their stellar masses during violent star formation episodes (Hammer et al. 2005), and half of them are transformed from unstable kinematics and peculiar morphologies to regular, relaxed galaxies dominated by thin disks. This suggests that galaxy collisions and their reminiscence may play a major role during the elaboration of their disks. Another line of support for such a suggestion is provided by Rodrigues et al. (2008, see also Rodrigues et al., this volume). The evolution of chemical abundance of the gaseous phases of galaxies shows a linear slope from z=0 to z=3, in strong contradiction with close box models that predict a much moderate evolution at z=0.4-0.8. Thus galaxies are not isolated systems and are exchanging gas, as expected during interactions.

\section{The elaboration of disk galaxies and of the Hubble sequence: disk rebuilt after mergers?}

There is a considerably growing set of evidences that the elaboration of disks is linked with galaxy mergers. The spiral rebuilding scenario was proposed by Hammer et al. (2005) to explain the observations of distant galaxies, including the simultaneous evolution of the global stellar mass, Luminosity-Metallicity relationship, pair statistics, IR light density, colors of spiral cores and number density of peculiar galaxies. This is supported by a similar evolution of their kinematics.  Since Barnes et al. (2002), simulations have  shown that under the condition of enough large gas content (generally $f_{gas}$ $\ge$ 50\%), major mergers may lead to the formation of a new disk (Robertson et al., 2006; Governato et al., 2007; Hopkins et al., 2008). Such high gas fractions are currently observed in the distant Universe at z$\sim$ 2 (Erb et al. 2007) and even at z$\sim$ 1 (Liang et al. 2006; Rodrigues et al. 2008).

Is the high fraction of anomalous kinematics consistent with a merger hypothesis? The most robust quantity derived to estimate the merger rate is the pair fraction of intermediate mass galaxies at z$\sim$ 0.6, for which all studies find 5$\pm$1\% (see e.g. Bell et al., 2006 and more recently, Lotz et al. 2008; Rawat et al. 2008). At first glance this appears to be contradictory with the higher fraction (26\%) of galaxies with complex kinematics. However the latter are found in pairs or could be merger remnants, so they could be quite numerous. Galaxy simulations predict a relatively small time scale for pairs to merge($\tau_{pair}$=0.35-0.5 Gyrs for M* galaxies). There are 5 times more galaxies with complex kinematics than galaxies in pairs: this is consistent if the complex velocity fields are related to merging, either during the first interaction in the pair or during the remnant phase. Such a reminiscence phase duration would be:  $\tau_{remnant}$= 3-5 $\times$ $\tau_{pair}$. Such values ($\tau_{remnant}$= 1.2-2 Gyrs) are indeed predicted by simulations of gas rich mergers to rebuilt significant disks (e.g. Robertson et al., 2006; Governato et al., 2007; Hopkins et al., 2008). 

Is the merger hypothesis consistent with the evolution of Tully Fisher relation and angular momentum? Puech et al. (2007a) have shown that galaxies with anomalous kinematics are outliers in the Tully Fisher and the $j_{disk}$-$V_{rot}$ relations. They demonstrated that major mergers can easily explain such a deviation during which galaxies are experiencing a random walk evolution in such planes.

How this could be reconciled with the past history of the Milky Way? The Milky Way disk is well known to have not been impacted by significant collisions since z=3-4. However, many evidences show that other spirals in the same mass range (e.g. M31, M81) have had a much more tumultuous past history (Ibata et al. 2005; Brown et al., 2008 and in this volume, Davidge, 2008). By comparing Milky Way properties to those of other spirals from the SDSS, Hammer et al. (2007) have shown that the Milky Way is rather exceptional, having a too small radius, angular momentum and stellar mass. Only 7$\pm$1\% of local spirals are Milky Way-like, conversely to M31 which is a rather ordinary spiral. Milky Way could be even more exceptional: it is still the only known galaxy in its mass range with an essentially primordial halo. 

\begin{figure}[t]
\begin{center}
\includegraphics[scale = 0.5]{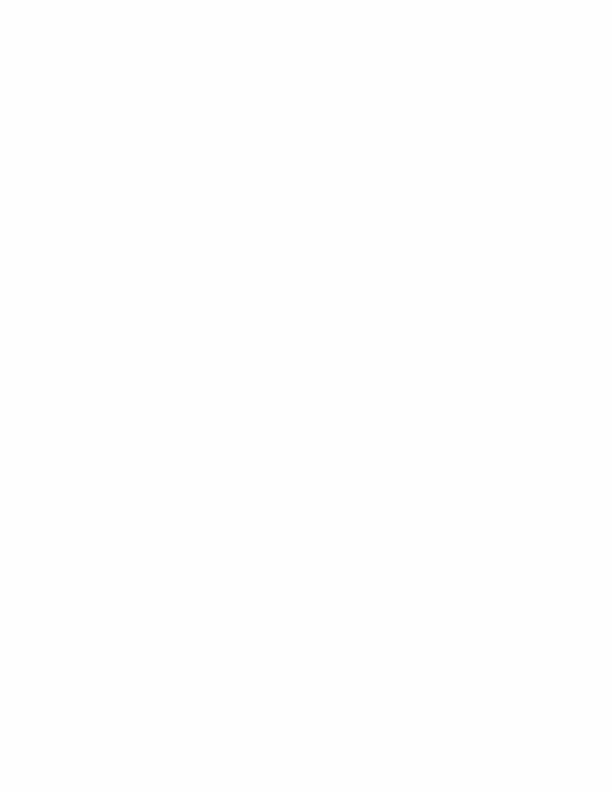}
\end{center}
\vspace*{-0.8cm}
\caption{Morphologies and kinematics (same symbols than in Figure 1) for two distant galaxies. On the left, J033239.72-275154.7 at z=0.4158 (modelled with GADGET2 including velocity field in the bottom); on the right, J033245.11-274724.0 at z=0.43462 (bottom: modelled with ZONE from Barnes et al, 2002, with dispersions seen in red colors). }\label{fig1}
\end{figure}

\section{Modelling z $\sim$ 0.6 galaxies with a similar accuracy than for local galaxies and concluding remarks}

Whether galactic disks have been produced as a by product of the last major merger may considerable change the theory of galaxy formation by superseding the tidal torque theory. Hopkins et al. (2008) have demonstrated that most of the process in the disk rebuilding phase is fundamentally dynamical, while feedback mostly remove part of the gas at large radii and preserve the disk from fragmentation. A crucial test is to examine with sufficient details the real galaxies and derive their accurate properties.

At z$\sim$ 0.6, the IMAGES galaxies sample an epoch of decreasing star formation, although significantly higher than at present epoch. To examine them we have a huge amount of details provided by the combination of their kinematics at $\sim$ 3kpc scales, their morphologies at $\sim$ 200 pc scales, their star formation rate (from Spitzer/MIPS photometry), and their stellar population properties (from their VLT/FORS2 spectra). For example Puech et al. (2007b) have been able to identify the impact of a 1:18 satellite infall in J033226.23-274222.8 at z= 0.66713 (see Figure 1, third panel), which is reponsible for only a small fraction of the star formation activity in this galaxy. The modelling of z$\sim$ 0.6 galaxies may be done following the steps of former models of nearby systems. The giant and star-bursting bar as well as the complex kinematics of J033239.72-275154.7 (see Figure 2) has been modelled by Peirani et al. (2008) by a 1:3 merger, the resulting galaxy being an S0$_a$. Yang et al. (2009, submitted) has shown that for J033210.76-274234.6 (see Figure 1, fifth panel), the 1:6 satellite has impacted the main galaxy with a very inclined orbit and near its core, letting the disk survive to the collision. Hammer et al. (2009) show that a compact LIRG (J033245.11-274724.0 at z=0.43462, see Figure 2) may have all its very peculiar properties (disk redder than the bulge, dynamical axis well offset from the optical main axis) reproduced by a 1:1 gaseous rich merger remnant leading to a Sc galaxy. Puech et al. (2008b, see arXiv0811.3893) has identified that for a significant part of J033241.88-274853.9, a gas rich galaxy at z=0.66702, the gas is ionised by shocks, revealed to large gas dispersions, which is predicted by a model of a 1:1 gas rich merger remnant.

More detailed analyses are in progress. Having been done for a representative mass selected sample of galaxies at z$\sim$ 0.6, they will definitively probe whether the present-day Hubble sequence has been elaborated from the last major merger, from ellipticals to late type spirals. Comparison with state of the art simulations would be invaluable for understanding the detailed paths of angular momentum built-up, the influence of gas viscosity, the way the expelled material can be re-accreted during the post-merger stage and possibly the formation of bars, rings and other spiral disk features. A detailed examination of the physics of the galaxy formation is now within our range: we may revisit how the Hubble sequence has been elaborated as well as how disks have acquired their large angular momentum and thin disks.

{\small
\acknowledgements 
I express my gratefulness to the organisers, the administrative persons and students for organising such an excellent meeting.
}





\end{document}